\documentclass[12pt]{article}
\input xy

\xyoption{all}

\usepackage[dvips]{graphicx}
\usepackage{amsfonts,amssymb,latexsym}

\def\s2{\frac{1}{\sqrt2}}

\def\be{\begin{equation}}
\def\ee{\end{equation}}
\def\beqa{\begin{eqnarray}}
\def\eeqa{\end{eqnarray}}

\def\Dsl{\,\raise.15ex\hbox{/}\mkern-13.5mu D} 
\def\d3{d^3}

\def\IR{\mathbb{R}}

\newcommand{\eq}{\begin{equation}}
\newcommand{\eqn}{\end{equation}}

\newcommand{\vep}{{\cal E}}

\newcommand{\half}{\frac{1}{2}}
\newcommand{\halfi}{\frac{i}{2}}

\def\IR{\relax{\rm I\kern-.18em R}}
\begin{document}

\vspace{.15cm}
\begin{center}
{\large {\bf Renormalization Group Flow for Noncommutative \\
Fermi Liquids}} \vspace{1cm}

{\large Sendic Estrada-Jim\'enez$^{a}$\footnote{e-mail: {\tt
sestrada@unach.mx}}, Hugo
Garc\'{\i}a-Compe\'an$^{b}$\footnote{e-mail: {\tt
compean@fis.cinvestav.mx}}} \\
and
{\large Yong-Shi Wu$^{c,d}$\footnote{e-mail: {\tt wu@physics.utah.edu}}}\\

{\small \em $^a$ Centro de Estudios  en F\'{\i}sica y
Matem\'aticas B\'asicas y Aplicadas \\
 Universidad Aut\'onoma de Chiapas,
Calle 4$^a$ Oriente Norte 1428 \\
Tuxtla Guti\'errez, Chiapas, M\'exico}
\\

{\small \em $^b$ Departamento de F\'{\i}sica, Centro de Investigaci\'on y
de Estudios Avanzados del IPN}\\
{\small\em P.O. Box 14-740, 07000 M\'exico D.F., M\'exico}\\

{\small \em $^c$ Department of Physics and Astronomy, University of Utah}\\
{\small \em Salt Lake City, UT 84112, USA} \\

{\small \em $^d$ Department of Physics, Fudan University, Shanghai
200433, China}

\vspace*{1cm}
\small{\bf Abstract} \\
\end{center}

\begin{center}
\begin{minipage}[h]{14.0cm} {Some recent studies of the AdS/CFT
correspondence for condensed matter systems involve the Fermi
liquid theory as a boundary field theory. Adding $B$-flux to the
boundary D-branes leads in a certain limit to the noncommutative
Fermi liquid, which calls for a field theory description of its
critical behavior. As a preliminary step to more general
consideration, the modification of the Landau's Fermi liquid
theory due to noncommutativity of spatial coordinates is studied
in this paper. We carry out the renormalization of interactions at
tree level and one loop in a weakly coupled fermion system in two
spatial dimensions. Channels ZS, ZS' and BCS are discussed in
detail. It is shown that while the Gaussian fixed point remains
unchanged, the BCS instability is modified due to the space
non-commutativity.}
\end{minipage}
\end{center}

\bigskip

\date{\today}

\vspace{0.5cm}

\leftline{April 2011}

\newpage

\section{Introduction}

Since more than ten years ago, noncommutative quantum field theory
arising from string theory \cite{hw,li,cds,sw} has received a
great deal of attention. (For excellent reviews, see
\cite{douglas,szabo}. For a reprint volume, see \cite{wubook}. The
Wightman axioms for noncommutative quantum field theory were
studied in \cite{alvarez-gaume}.) One of the most interesting
results in the analysis of perturbative dynamics of noncommutative
scalar field theories on a Euclidean space $\mathbb{R}^4$, is the
existence of a mixing of the UV and IR behaviors \cite{minwalla}
and its origin from the theory of open strings \cite{sw}.

However application of noncommutative quantum field theory to low
energy (low temperature) many-body systems other than the
fractional quantum Hall effect (FQHE) has not been yet discussed
extensively in the literature. It is known that the low energy
dynamics of a quantum many-body system can be described by an
effective field theory. It would be interesting to study the
modifications due to spatial noncommutativity. In this paper we
are particularly interested in studying the effective field theory
of non-relativistic weakly interacting fermions at low
temperature.

In the absence of spatial noncommutativity, such systems are
described phenomenologically, as well as quantum field
theoretically, by the so-called {\it normal Fermi liquid theory}
(see Refs. \cite{nozieres,abrikosov,pines,landau} for a
traditional perspective). This theory has also been studied in the
functional integration approach in \cite{popov}. Recently, a great
deal has been worked out concerning the Fermi liquid theory in the
context of an effective field theory and its characterization in
terms of the renormalization group
\cite{gallavotti,polchinski,shankar,frolich,dupuis,belitz,wen}.
Within this modern perspective the renormalization group methods
have been used to study the interacting fermion systems, and the
Landau's theory of the Fermi liquid is derived as a fixed point of
the renormalization group flow.  The Landau's theory of Fermi
liquid theory is a very important paradigm: it may be implicitly
"hidden" in unphysical Hilbert spaces, as suggested in
\cite{anderson}, in phenomena like high $T_c$ superconductivity
and the FQHE, which are normally thought of as non-Fermi liquids.

Recently the string/M theory community has also shown great
interest in the theory of Fermi liquids, and has been able to
relate it to various situations and processes. For instance it was
found in \cite{Horava:2005tt} that non-critical M-theory in $2+1$
dimensions can be described in terms of a non-relativistic Fermi
liquid. Moreover, the topology of the Fermi surface in the Fermi
liquid theory has also been described through K-theory
\cite{horava}. In the context of the AdS/CFT correspondence there
are also certain relations with the Fermi liquid theory.
Semiconductors also have been studied in this context predicting
the dynamical generation of mass gap and metal-insulator quantum
phase transition at zero temperature \cite{Rey:2008zz}. Moreover
it has been shown that string theory in the background of dyon
black holes in four-dimensional anti-de Sitter spacetime is
holographic dual to conformally invariant composite Dirac fermion
metal described by a relativistic Fermi liquid theory
\cite{Bak:2009kz}. Non-Fermi liquids are also studied from the
same perspective \cite{Lee:2008xf}. Detailed analysis of
computations of the correspondence implies the existence of a
Fermi liquid operators in $N = 4$ SYM whose anomalous dimensions
behave similar to Fermi liquids in condensed matter systems
\cite{Berkooz:2008gc}. Further analysis of the gravitational dual
of the Fermi liquid in the $N=4$ super Yang-Mills theory coupled
to fundamental hypermultiplet at nonvanishing chemical potential
have been studied in \cite{Kulaxizi:2008kv}. Here an interesting
checking on the structure of the zero sound and the first sounds
was verified. More recently, in Ref. \cite{Sachdev:2010um}, it is
was found an interesting description of the physical properties of
holographic metals near charged black holes in anti-de Sitter
space, and the fractionalized Fermi liquid phase of the lattice
Anderson model.

In the present work, we introduce noncommutativity between spatial
coordinates, in order to avoid unitarity and causality problems.
Namely, the noncommutativity will be defined in
$\mathbb{R}^d_\star \times \mathbb{R}$: We have $d$ noncommutative
spatial coordinates that satisfy
$[\widehat{x}^i,\widehat{x}^j]=i\Theta^{ij},$ where $\Theta^{ij}$
is antisymmetric and real. The time $t$ is considered as a
commutative coordinate, so we require that $\Theta^{0i}=0.$ For
the study of the field theory in this space one usually think of a
deformation in the product in the space of functions, i.e., the
noncommutative space $\mathbb{R}_\star^d$ can be regarded as the
algebra over the usual $\mathbb{R}^d$ with a deformation of the
product of functions into the Moyal star product, defined by \eq
(\phi_1 \star \phi_2)(x) = \exp \big({{i \over 2} \Theta^{\mu \nu}
\partial_\mu^y
\partial_\nu^z}\big) \phi_1(y)\phi_2(z) |_{y=z=x}.
\eqn One property of this product is that the quadratic part of
the action in a field theory, up to a total divergence, is exactly
the same as that in the commutative case. Therefore the
propagators remain the usual ones, while the noncommutativity
modifies the interactions.

Some earlier papers on noncommutative field theory in the context
of condensed matter systems have been collected in the book
\cite{wubook}, mainly on the FQHE. For more papers see, for
instance, \cite{Read:1998dn,jackiw1,barbon,landsteiner}.
Renormalization group flow in noncommutative Landau-Ginzburg
theory for thermal phase transitions of Bose fluids was presented
in \cite{hong}. The aim of this paper is to discuss
renormalization group flow in the noncommutative Fermi liquid
theory. Recently some non-relativistic systems in noncommutative
spaces have been discussed in references
\cite{bak1,gomis2,mateos,bak2,toporder}. The study of the
renormalization group flow in the noncommutative Gross-Neveu model
of interacting fermions has been carried out in
\cite{Akhmedov:2001fd}. Moreover, a superconducting vortex liquid
system in the lowest Landau level approximation was studied from
the view point of the noncommutative field theory in
\cite{Yeo:2003tj}. We expect that our study of renormalization
group flow for noncommutative Fermi liquids could shed more light
on the above topics and subjects.

The present paper is organized as follows: in section 2 we give an
overview of the Fermi liquid theory in order to introduce the
notation and conventions. In section 3 we introduce the
noncommutative deformation of the Fermi liquid theory, in
particular, the tree-level renormalization is carried out. Section
4 is devoted to the study of the one-loop renormalization in two
spatial dimensions. Conclusions and final remarks are compiled in
section 5.

\section{Overview of the Normal Fermi Liquid Theory}

We will understand by a normal Fermi liquid a system of
non-relativistic weakly interacting fermions in, say, $3+1$
dimensions. The first description for this system was proposed by
Landau, and it is of a phenomenological nature \cite{landau}. The
main assumption of the Landau approach to Fermi liquid, is that
there exists a one to one correspondence between the electrons in
a non-interacting gas of fermions and some elementary excitations
of the interacting system known as {\it quasiparticles}. These
excitations are characterized by its energy $\vep(\bf p)$. Starting
from the assumption that the ground state of the weakly
interacting system can be generated adiabatically from some
eigenstate of the ideal system. The effective interactions will be
reflected in the behavior of the quasiparticles,  as effective
particles, i.e. free particles {\em dressed} with the interaction.

The quasiparticle picture makes sense only very near from the
Fermi surface, i.e. it is relevant only for excitations at very
low temperature.  It is well known that many materials behave as a
Fermi liquid at temperatures much below the Fermi energy. Moreover
if we consider a pure system at zero temperature, the life-time of
the quasiparticles changes as the inverse of the square of $E -
E_F$, where $E_F$ is the Fermi energy.

Let $n(\bf p)$ be the distribution function of quasiparticles, this
function must be defined in such a way that the energy of the
Fermi liquid $E$ is determined in a unique way and the ground
state corresponds to the distribution function in which all states
inside the Fermi surface are occupied. In the ideal system the
relation between the energy of any state and their corresponding
distribution function is given by \eq E=\int_{\bf p} n({\bf p})
\frac{{\bf p}^2}{2m} \ \frac{d^3p}{(2\pi\hbar)^3}. \eqn

Once the interaction is taken into account the last relation is
modified, now this can be expressed through a functional relation
$E[n({\bf p})]$, whose form depend of the distribution of all
particles in the liquid, and in general, we cannot know it
explicitly. Nevertheless if  $n({\bf p})$ is very close to the
distribution function of the ground state it is convenient make a
Taylor expansion of $E[n({\bf p})]$ around this state and we get to the
first order

\eq E = E_0 + \int_{\bf p} {\cal E}({\bf p}) \delta n({\bf p})  \
\frac{d^3p}{(2\pi\hbar)^3}, \eqn where $E_0$ is a fixed ground state
energy and ${\cal E}({\bf p})$ is the functional derivative of $E$ with
respect to distribution function and it is also a functional of
$n({\bf p})$. If $\delta n({\bf p})$ describes a state with an additional
quasiparticle with momenta ${\bf p}$, the energy of this state is $E_0+
\int_{\bf p} \vep({\bf p})d^3p/(2 \pi \hbar)^3 $, then  $\vep({\bf p})$ is
related to the energy of the quasiparticle. On the Fermi surface
$\vep({\bf p})$ is associated to the Fermi energy $\vep_F$ which at
zero temperature corresponds to a chemical potential $\mu |_{T=0} =
\vep_F = \vep({\bf p}_F).$

Near the Fermi surface we can expand $\vep({\bf p})$ around it
as
\eq
 {\cal E}({\bf p})=\vep_F +
{\bf v}_{F}\cdot({\bf p} - {\bf p}_F) + \cdots \label{velocidad} \eqn where
${\bf v}_F = \nabla_{\bf p} {\cal E}\big|_{{\bf p}={\bf p}_F}.$ In the case when
these excitations correspond to real particles of the system we have
that ${\bf v}_F={\bf p}_F/m$.

\subsubsection{Interaction of Quasiparticles}

We remark that when the distribution function is changed, for
instance, by adding a quasiparticle, it changes not only the total
energy of the system, but also it changes the energy of the
quasiparticles $\vep$. This is because $\vep$ is a functional of
the density. Since the total energy of the system is not the
simple sum of the individual energy of each quasiparticle, it is
necessary to consider an expansion to second order as follows

\eq E-E_0=\int_{\bf p} (\vep({\bf p}) - \mu )\delta n({\bf p}) +
\frac{1}{2}\int_{{\bf p},{\bf p}^\prime} f({\bf p},{\bf p}^\prime)\delta n({\bf p})
\delta n({\bf p}^\prime), \eqn where the sub-indices of the integrals
stand for the integration variables. The coefficient
$f({\bf p},{\bf p}^\prime)$ is the second functional derivative of the
energy respect to density functional and is known as the
interaction term of the quasiparticles. (It vanishes for the ideal
Fermi gas.) For low energy excitations, the variations $\delta
n({\bf p})$ and $\delta n({\bf p}^\prime)$ for the two quasiparticles are
non-zero only for ${\bf p}$ and ${\bf p}^{\prime}$ near the Fermi surface.
For this reason the function $f({\bf p},{\bf p}^\prime)$ is, in practice,
evaluated only on the Fermi surface $|{\bf p}|=|{\bf p}^\prime|=|{\bf p}_F|$, so
it depends only on the directions of ${\bf p}$ and ${\bf p'}$ and the spin
$\sigma$ and $\sigma'$, respectively, of the quasiparticles.

In the Landau's theory, the deviation from equilibrium state of
the Fermi liquid is studied through the Boltzmann transport
equation, with the usual conditions, that the de Broglie wave
length of the quasiparticle must be small in comparison with the
characteristic wave length where the distribution function varies
considerably. Furthermore we can see that the collision of the
quasiparticles produces ordinary hydrodynamic sound waves. However
it is also found that when the system is at zero temperature there
must exist another type of ``sound waves", to which the collision
of quasiparticles is not relevant. What is relevant is the change
in shape of the Fermi surface at different spacetime points. This
sound waves are known as {\em zero sound waves} (ZS).

Fermi liquid theory has been studied also from the view point of
the quantum field theory by using the canonical formalism,
recovering the Landau's Fermi liquid theory in the normal phase as
well as in the superfluid phase \cite{abrikosov,landau}. In this
formalism the four-point proper vertex is related with the zero
sound waves. This part of the vertex function is called zero sound
channel (ZS), and this is the most important process to recover
the phenomenological theory of the Fermi liquid.

From an effective theory perspective Polchinski \cite{polchinski}
was able to recover the Landau's theory of Fermi liquid. In this
work a system of interacting fermions was studied from the
symmetries, respected by the possible terms in the action of the
system. In this context we are interested in the computation of
the possible modifications that arise in a theory of interacting
fermions in noncommutative space.

\section{Noncommutative Fermi Liquid Theory}

Though normally space noncommmutativity is considered as
originated from small distance physics, it is well known to lead
non-local properties such as the UV/IR mixing. We are interested
in a non-relativistic effective theory, which is defined with a UV
cutoff $\Lambda$ such that the degrees of freedom with
$p=|{\bf p}|>\Lambda$ do not enter the description of the system. It
is reasonable to study the nontrivial effects induced by
noncommutativity on large distance physics. In favor of this idea,
we note that working at low energies do not prevent the emergence
of new characteristics in noncommutative theories \cite{gomis2},
because of the UV/IR mixing. From the view point of the particles,
we may also argue that at low energies, the effective interactions
between particles are not genuinely point-like. Moreover, as
mentioned in the book \cite{wen}, the electrons can be considered
as effectively non-local particles due to their Fermi statistics.
Therefore, it is legitimate to examine non-local effects arising
from space noncommutativity on the interactions in the effective
theory.

\subsection{The Effective Action}

In this subsection we briefly overview the effective action. We
adopt the euclidean formulation of the functional integral
formalism. For a system of interacting fermions the partition
function is given by

\eq Z[\eta, \eta^\dagger]=\int [d\psi][d\psi^\dagger] \exp
\bigg(-S_0-S_I+ \int \psi^\dagger \eta + \int \eta^\dagger
\psi\bigg), \eqn where $\eta$ and  $\eta^\dagger$ are the sources,
$\psi$ and $\psi^\dagger$ are the fermionic fields which we take
as Grassmann variables. $S_0$ is the free action and $S_I$ is the
interacting term.

For the usual Fermi gas the two-point function is given by

\begin{equation}
G({\bf x},{\bf y})= \int \frac{d^{d+1}P}{ (2\pi)^{d+1}} \frac {\exp\bigg\{
i\bigg[ {\bf P} \cdot ({\bf x}-{\bf y}) +
 i\omega(\tau_x-\tau_y)\bigg]\bigg\}}{ i\omega -
\bigg(\frac{{\bf P}^2}{2m}-\mu \bigg)}.
\end{equation}

For a liquid of fermions, it is usual to assume that a substantial
change occurs, in the weakly interacting theory, in the propagator
$G({\bf x},{\bf y})$, which is of the form:
\eq G({\bf x},{\bf y}) = \int
\frac{d^{d+1}P }{ (2\pi)^{d+1}} \frac {\exp\bigg\{ i\bigg[
{\bf P} \cdot ({\bf x}-{\bf y}) + i\omega(\tau_x-\tau_y)\bigg]\bigg\}} {
i\omega - \left({\cal E}({\bf P})-{\cal E}_F\right)}. \eqn

Since we are interested only in correlations at low energy, let us
define two sets of variables \cite{polchinski,shankar,hong}:
$$
\phi_< = \phi ({\bf P})\qquad {\rm for}\qquad 0<P<\Lambda/s,
$$
\eq \phi_>=\phi({\bf P})\qquad {\rm for}\qquad \Lambda/s \leq P \leq
\Lambda. \eqn Our action can be divided into two parts,
corresponding to fast modes $\phi_>$ and slow modes $\phi_<$, so
that the total action is given by \eq S[\phi_<,\phi_>] =
S_0(\phi_<) + S_0(\phi_>) + S_I(\phi_<,\phi_>). \eqn Then the
partition function is given by \eq Z= \int [d \phi_>]
[d\phi_<]e^{-S_0(\phi_<)} e^{-S_0(\phi_>)}
e^{-S_I(\phi_<,\phi_>)}, \eqn which can be rewritten as: \eq
Z=\int [d\phi_<] e^{-S'(\phi_<)}. \eqn This defines the effective
action $S'(\phi_<)$: \eq e^{-S'(\phi_<)}= e^{ -S_0(\phi_<)} \int
[d \phi_>] e^{-S_0(\phi_>)} e^{-S_I(\phi_<,\phi_>)}. \eqn This
expression can be further rewritten as \eq e^{-S'(\phi_<)}=e^{
-S_0(\phi_<)} \bigg\langle e^{-S_I(\phi_<,\phi_>)}
\bigg\rangle_{0>}, \eqn where $\langle \cdot \rangle_{0>}$ stands
for the average value with respect to the fast modes of the action
$S_0$. This effective action can be computed through approximation
methods by means of the cummulant expansion, that relates the
correlation function of the exponential with the exponential of
the correlations functions, i.e. \eq \big\langle e^{\Omega}
\big\rangle = \exp\bigg\{ \langle \Omega \rangle + \half [\langle
\Omega^2 \rangle - \langle \Omega \rangle^2] + \cdots \bigg\}.
\eqn

We construct an effective action which defines new coupling
functions between the fields. These new functions must be compared
with the original ones in the action, but these quantities are
defined in different kinematic regions, $0<P<\Lambda/s$ and
$0<P<\Lambda$, respectively. So it is necessary to rescale the
momenta ${\bf P}'=s {\bf P}$ in the effective action to recover the
original scale. We also need to rescale the fields to define the
new fields: \eq \phi'({\bf P'})= \zeta^{-1}\phi_<({\bf P'}/s), \eqn where
we choose the real prefactor $\zeta$ so that the quadratic part of
the action in terms of the new fields have a fixed coefficient
(independent of $s$).

In summary, the renormalization process goes in three steps: 1)
Eliminate the fast modes, that is to integrate out the momenta
with values inside the interval $[\Lambda/s,\Lambda]$; 2)
Introduce a momentum scaling ${\bf P}\rightarrow s {\bf P}$ and recover
the original cutoff $\Lambda$; 3) Introduce the scaled fields
$\phi'({\bf P'})=\zeta^{-1}\phi_<({\bf P'}/s)$ and rewrite the effective
action in terms of the new fields. The quadratic kinetic term of
the action should have the same coefficient as before.

In practice, one carry out the above renormalization procedure in
two stages: First we look at the free action and fix the
coefficient by appropriate rescaling of the new fields. In this
way, we will be able to find the Gaussian fixed point,
corresponding to an ideal non-interacting system. After that we
examine how the interaction terms scale under renormalization, and
classify them as relevant, irrelevant or marginal terms under the
renormalization group flow.

In particular for our theory, the free action is given by \eq
S=\int_{-\infty}^\infty d\omega \int_{-\Lambda}^\Lambda d^dP
\bigg( \psi^\dagger_\sigma({\bf P})i\omega \psi_\sigma({\bf P}) -
({\cal E}({\bf P})-{\cal E}_F )\psi^\dagger_\sigma({\bf P}) \psi_\sigma({\bf P})\bigg),
\eqn where $\sigma$ is the spin index and ${\cal E}_F$ is the Fermi
energy that correspond to chemical potential at zero temperature.

The first step is to integrate out the fields $\psi$ and
$\psi^\dagger$ in the partition function within
$\Lambda/s<P<\Lambda$, which means integrating out the fast modes.
We can see that this results in a Gaussian integral, up to an
irrelevant numerical factor.

In view of the facts that the ground state is determined by the
Fermi surface, and that when the energy goes to zero the momentum
must go to the Fermi surface, it is natural to write the momentum
of our excitation as \eq {\bf P}={\bf k}+{\bf p}, \eqn where ${\bf k}$ is a vector
on the Fermi surface and ${\bf p}$ is normal to this surface.

As we are interested only in the region near to Fermi surface, the
generic energy ${\cal E}$ can be expanded in series as \eq {\cal
E}({\bf P}) -{\cal E}_F = {\bf p} \cdot {\bf v}_F({\bf k}) +
O({\bf p}^2)\label{expan1}. \eqn With this decomposition, we should
scale the momentum as ${\bf k} \rightarrow {\bf k}$, ${\bf p} \rightarrow s
{\bf p}$ and $\omega \rightarrow s\omega$. Making the substitution in
the free action we find that the field scales as $s^{-3/2}$, i.e.
$\psi'(\omega,{\bf k}',{\bf p}')=s^{-3/2}\psi(\omega,{\bf k},{\bf p})$.

We will focus on studying the Fermi liquid in a {\em two}
dimensional plane (i.e. $d=2$) with a {\em circular} Fermi
surface. For this case the momentum decomposition (\ref{expan1})
is still valid, but note that \eq |{\bf p}|=|{\bf P}|-|{\bf k}|. \eqn Then,
the integral measure in polar coordinates is $P_F\,dp\,d\phi,$
where $P_F= |{\bf k}|$. As we are interested only in the region close
the Fermi surface, we only need to scale the radial component
according the previous prescription. The free action becomes \eq
S_0=\int \, \frac{d \omega}{2\pi} \int \, \frac{d\phi}{2\pi} \int
\, \frac{dp}{2\pi} \psi_\sigma^\dagger (\omega,\phi,p)(i\omega -
pv_F) \psi_\sigma(\omega,\phi,p), \eqn where we replaced the
measure $Pdp$ by $P_Fdp$ and absorbed a factor of $\sqrt{P_F}$ in
each one of the fermionic fields. With this scaling for the fields
the free action is a fixed-point action, and we can make the
perturbative expansion around it.

\subsection{Proper Form of the Noncommutative Interaction Term}

The noncommutative interaction action that we study is the quartic
interaction term, which in the coordinate representation is given
by \eq \label{quarticinterc}S_{I}=\int d\tau \, d^2x \, d^2y \, \psi^\dagger({\bf x}) \star
\psi({\bf x})  \star V({\bf x} -  {\bf y}) \star \psi^\dagger({\bf y}) \star
\psi({\bf y}). \eqn We note that we consider only noncommutativity
between spatial coordinates, and our potential is time
independent. So we can write this integral as \eq S_{I} = \int
d^2x\,d\tau_x \, d^2y \, d\tau_y \, \psi^\dagger({\bf x}) \star
\psi({\bf x}) \star V({\bf x} - {\bf y})\delta(\tau_x-\tau_y) \star
\psi^\dagger({\bf y}) \star \psi({\bf y})\label{interac1}. \eqn

To make considerations of symmetry constraints simpler, we will
work in momentum space. The above expression can be rewritten as
\eq S_I = \int_{\bf P} \psi^\dagger({\bf P}_4) \psi({\bf P}_3)
\psi^\dagger({\bf P}_2) \psi({\bf P}_1) V({\bf P}_4,{\bf P}_3,{\bf P}_2,{\bf P}_1)
e^{-\frac{i }{ 2}({\bf P}_1 \wedge {\bf P}_2 + {\bf P}_3\wedge {\bf P}_4)},
\label{interac2} \eqn where we write explicitly the star product
as ${\bf p} \wedge {\bf q} \equiv \Theta^{\mu \nu} p_\mu q_\nu$.


Before studying the scaling of the fields to classify the
interaction potential, we need to check the behavior of our action
under the interchange of particles. Re-ordering the terms in the
integral (\ref{interac2}), considering the rules of the Grassmann
variables and renaming of the variables, we have \eq S_I =
\int_{\bf P} \psi^\dagger({\bf P}_4) \psi^\dagger({\bf P}_3) \psi({\bf P}_2)
\psi({\bf P}_1) V({\bf P}_4,{\bf P}_3,{\bf P}_2,{\bf P}_1) e^{\big[-{i \over 2}({\bf P}_1
\wedge {\bf P}_4 + {\bf P}_2\wedge {\bf P}_3)\big]}. \eqn

For the usual commutative case it is necessary to impose that the
interaction potential be antisymmetric with respect to their
variables, in such way that the action is invariant under the
change of the order of the fields, i.e., $V({\bf P}_4, {\bf P}_3,
{\bf P}_2,{\bf P}_1) = V({\bf P}_3, {\bf P}_4, {\bf P}_1,{\bf P}_2)= -
V({\bf P}_3,{\bf P}_4,{\bf P}_2,{\bf P}_1) = - V({\bf P}_4, {\bf P}_3, {\bf P}_1,{\bf P}_2)$.
However for our present case, we have an additional phase factor
coming from space noncommutativity. The presence of this factor
makes the symmetry consideration in the noncommutative case a bit
more complicated.

If we exchange the labels of momenta ${\bf P}_4$ and ${\bf P}_3$, the
integral must be unchanged: \eq S_I = \int_{\bf P} \psi^\dagger({\bf P}_3)
\psi^\dagger({\bf P}_4) \psi({\bf P}_2) \psi({\bf P}_1)
V({\bf P}_3,{\bf P}_4,{\bf P}_2,{\bf P}_1) e^{-\halfi({\bf P}_1 \wedge {\bf P}_3 +
{\bf P}_2\wedge {\bf P}_4)}. \eqn Now, interchanging the fields with
momentum labels ${\bf P}_3$ and ${\bf P}_4$, we need to introduce a minus
sign as follows
\eq S_I= -\int_{\bf P} \psi^\dagger({\bf P}_4)
\psi^\dagger({\bf P}_3) \psi({\bf P}_2) \psi({\bf P}_1)
V({\bf P}_3,{\bf P}_4,{\bf P}_2,{\bf P}_1) e^{ -\halfi({\bf P}_1 \wedge {\bf P}_3 +
{\bf P}_2\wedge {\bf P}_4)}.
\eqn
Moreover, we can absorb the minus sign
using the antisymmetry property of the interaction potential and,
for the usual case, we recover the original action. But in this
process we also have an additional phase factor, which is not the
same as before, then we need to add the two integrals in order to
recover the symmetry of the action. Thus, we are finally led to
the action given by
\eq S_I= \int_{\bf P} \psi^\dagger({\bf P}_4)
\psi^\dagger({\bf P}_3)\psi({\bf P}_2)\psi({\bf P}_1)
U_\Theta({\bf P}_4,{\bf P}_3,{\bf P}_2,{\bf P}_1),
\label{primeraec}
\eqn
with \eq
U_\Theta({\bf P}_4,{\bf P}_3,{\bf P}_2,{\bf P}_1) = \half
V({\bf P}_4,{\bf P}_3,{\bf P}_2,{\bf P}_1) \left[ e^{-\halfi({\bf P}_1 \wedge {\bf P}_4 +
{\bf P}_2\wedge {\bf P}_3)} + e^{-\halfi({\bf P}_1 \wedge {\bf P}_3 + {\bf P}_2\wedge
{\bf P}_4)} \right].
\eqn
We have checked that with the additional
phase term in $U_{\Theta}$, the above action $S_I$ has the desired
antisymmetry property.

As particular example we can simplify the bilocal potential $V({\bf x}-{\bf y})$ by taking it to be a local interaction coupling constant by assuming $V({\bf x}-{\bf y})=g \delta({\bf x}-{\bf y})$ and substituting it in eq. (\ref{quarticinterc}) we get
\eq
S_{I}=g\int d\tau \, d^2x \, d^2y \, \psi^\dagger({\bf x}) \star
\psi({\bf x})  \star \delta({\bf x}-{\bf y}) \star \psi^\dagger({\bf y}) \star
\psi({\bf y}). \eqn
In momentum space we have a similar expression to the previous one for the general case. After reordering the fields we have
\eq S_I =
g\int_{\bf P} \psi^\dagger({\bf P}_4) \psi^\dagger({\bf P}_3) \psi({\bf P}_2)
\psi({\bf P}_1) e^{-{i \over 2}({\bf P}_1
\wedge {\bf P}_4 + {\bf P}_2\wedge {\bf P}_3)}. \eqn
if we make the same procedure  as earlier, and impose that the interaction coupling should keep the symmetries of the full vertex with four external lines as in general case, we find that this interaction term becomes
\eq S_I =
-g\int_{\bf P} \psi^\dagger({\bf P}_4) \psi^\dagger({\bf P}_3) \psi({\bf P}_2)
\psi({\bf P}_1) e^{-{i \over 2}({\bf P}_1
\wedge {\bf P}_3 + {\bf P}_2\wedge {\bf P}_4)}.
\label{newinteract}
\eqn
Then for compliance the requirements mentioned, we
can see that it is necessary to introduce two terms in the interaction term of the action, as has been proposed in \cite{castorina}  as a generalization of the noncommutative interaction Lagrangian for fermions,  in analogy the case of the complex scalar fields \cite{otro} and we have
\eq S_I =
\frac{1}{2}g\int_{\bf P} \psi^\dagger({\bf P}_4) \psi^\dagger({\bf P}_3) \psi({\bf P}_2)
\psi({\bf P}_1) \left[e^{-{i \over 2}({\bf P}_1
\wedge {\bf P}_4 + {\bf P}_2\wedge {\bf P}_3)}- e^{-{i \over 2}({\bf P}_1
\wedge {\bf P}_3 + {\bf P}_2\wedge {\bf P}_4)}\right], \eqn
where the minus sign between the phase terms is characteristic of this particular case.
Because this term vanishes in the commutative case, as it should be, when there are no internal degrees of freedom
(as here we are discussing spinless fermions) \cite{zinn}.
Finally we have seen that only with symmetry arguments of the Lagrangian, the additional term in the interaction action arise in a natural way.

Therefore space noncommutativity leads to the
appearance of additional phase terms that multiply the quartic
interaction. Following the renormalization group analysis
\cite{hong}, we keep the star product structure of this
interaction term intact, and apply renormalization group
transformations only to the coefficient function
$V({\bf P}_4,{\bf P}_3,{\bf P}_2,{\bf P}_1)$. Consequently as we will see in Sec. 4, the interactions of the type (\ref{newinteract}) are already included in the renormalization group flow from action (\ref{primeraec}).

The integral measure is given by
\eq \int_{\bf P} = \left[
\prod_{i=1}^3 \int_0^{2\pi}{d \phi_i \over 2\pi}
\int_{-\Lambda}^{\Lambda} {dp \over 2\pi} \int_{-\infty}^\infty {d
\omega \over 2\pi} \right]\theta(\Lambda-|p_4|), \label{medida}
\eqn where $p_4=|{\bf P}_4|-P_F$. In this measure we have
incorporated the constraints on the momenta due to energy and
momentum conservation. While energy conservation does not
constrain the integration over the remaining energy variable,
which can still take any value, the same is not true for the
momentum variables. The four momenta should be restricted to be in
a ring-shaped region of thickness $2\Lambda$ around the Fermi
surface. If we choose freely three of the momenta, the fourth
momentum could be outside this region. To avoid this situation, we
have introduced a Heaviside function for the fourth momentum.

\subsection{Renormalization of the Interaction at Tree Level}

Having obtained the complete form for the noncommutative
interaction term, we proceed to perform the renormalization group
analysis according to the procedure mentioned in the previous
subsection. First we note that as we have the step function in
(\ref{medida}) depending on $p_4$ we must study carefully the
scaling of this function, because $p_4$ depend not only on other
$p$'s but also on $P_F$ \cite{shankar}: \eq
p_4=\big|(P_F+p_1){\bf \Omega}_1+(P_F+p_2) {\bf \Omega}_2-
(P_F+p_3){\bf \Omega}_3 \big|-P_F, \label{p4}, \eqn where
${\bf \Omega}_i$ is the unit vector in the direction of ${\bf P}_i$,
i.e., ${\bf \Omega}_i= {\bf  i} \cos \phi_i + {\bf  j} \sin \phi_i.$
Here $\phi_i$ is the azimuthal angle of momentum $\bf P_i$.

Making the scaling of the momentum, we find that the step function
change as \eq \theta(\Lambda- |p_4|(p_1,p_2,p_3,P_F))\to
\theta(\Lambda- |p'_4|(p'_1,p'_2,p'_3,sP_F)). \eqn Therefore, the
step function $\theta$, after the renormalization group
transformation, does not have the same dependence on the new
variables as the $\theta$ function did before the transformation,
because $P_F \to s P_F$.

In order to understand how to scale the interaction part properly,
let us make a smooth cut  off for $p_4$: \eq
\theta(\Lambda-|p_4|)\to e^{-p_4/\Lambda}. \eqn We rewrite
(\ref{p4}) as
\[
p_4=\big|P_F({\bf \Omega}_1+{\bf \Omega}_2-{\bf \Omega}_3)+ p_1
{\bf \Omega}_1+ p_2 {\bf \Omega}_2- p_3 {\bf \Omega}_3 \big|- P_F,
\]
and define ${\bf \Omega}_1+ {\bf \Omega}_2- {\bf \Omega}_3= {\bf \Delta}$. In
the previous expression, we can drop the terms of order $O(\bf p)$,
because in the regime that we are interested this gives a sum of
order $\Lambda$ which will be smoothly suppressed by the
exponential decay and $p_4 \approx P_F$.

Under the renormalization group, at tree level we have
\begin{eqnarray}
&&\prod_{i=1}^3 \int_{-\Lambda}^\Lambda
\frac{dp_i}{2\pi}\int_0^{2\pi}\frac{d\phi_i}{2\pi}
\int_{-\infty}^\infty \frac{d\omega_i}{2\pi}
e^{-(P_F/\Lambda)||{\bf \Delta}|-1|}U_\Theta({\bf p},\omega,\phi)
\psi^\dagger \psi^\dagger \psi\psi \nonumber \\
&&\qquad \qquad \to \prod_{i=1}^3 \int_{-\Lambda}^\Lambda
\frac{dp'_i}{2\pi}\int_0^{2\pi}\frac{d\phi_i}{2\pi}
\int_{-\infty}^\infty \frac{d\omega'_i}{2\pi}
e^{-(sP_F/\Lambda)|{\bf\Delta}|-1|}
U_\Theta(\frac{\bf p'}{s},\frac{\omega'}{s},\phi)
\psi^\dagger \psi^\dagger \psi\psi. \nonumber \\
\end{eqnarray}

We write \eq e^{-(sP_F/\Lambda)||{\bf \Delta}|-1|}
=e^{-(P_F/\Lambda)||{\bf \Delta}|-1|}
e^{-[(s-1)P_F/\Lambda]||{\bf \Delta}|-1|}, \eqn so that the measure
after and before of the transformation have the same factor
$e^{-(P_F/\Lambda)||{\bf \Delta}|-1|}$. We can now compare the
actions and identify the new quartic coupling as
\eq
U_\Theta'({\bf p'},\omega',\phi)=e^{-[(s-1)P_F/\Lambda]|
|{\bf \Delta}|-1|}U_\Theta(\frac{\bf p'}{s},\frac{\omega'}{s},\phi).
\eqn

Thus, we conclude that the only coupling that survives the
renormalization group transformation without decay corresponds to
the cases in which \eq
|{\bf \Delta}|=|{\bf \Omega}_1+{\bf \Omega}_2-{\bf \Omega}_3|=1. \eqn Then
we can analyze the renormalizability properties focusing on the
cases that have a non-trivial contribution. Such cases are those
that satisfy the following angular conditions \cite{shankar}

\eq \label{cond1} {\rm Case \ I:} \qquad {\bf \Omega}_4= {\bf
\Omega}_1, \qquad ({\rm hence}\; {\bf \Omega}_2={\bf \Omega}_3)
\eqn \eq\label{cond2} {\rm Case \ II:} \qquad {\bf
\Omega}_4={\bf \Omega}_2, \qquad ({\rm hence}\;
{\bf\Omega}_1={\bf \Omega}_3) \eqn \eq\label{cond3} {\rm Case \
III:} \qquad {\bf \Omega}_1=-{\bf\Omega}_2, \qquad ({\rm
hence}\; {\bf \Omega}_3=-{\bf \Omega}_4). \eqn

Thus for couplings obeying these conditions, we have \eq
V'({\bf p'},\omega',\phi)=V({{\bf p'} \over s}, {\omega' \over s}, \phi).
\eqn It follows that the coupling function $V$ is renormalized to
a function that may depend on $\phi$ but independent of $\bf p$ and
$\omega$, when the cutoff is reduced (i.e. $s >1$).

We see that the tree-level fixed point is characterized by three
independent functions and not by a handful of couplings. They are
given by \eq U_\Theta[\phi_4=\phi_1;\phi_3=\phi_2;\phi_2;\phi_1] =
F_\Theta(\phi_1;\phi_2), \eqn \eq
U_\Theta[\phi_4=\phi_2;\phi_3=\phi_1;\phi_2;\phi_1] =
F_\Theta'(\phi_1;\phi_2), \eqn \eq
U_\Theta[\phi_4=\phi_3+\pi;\phi_2=\phi_1+\pi] =
V_\Theta(\phi_1;\phi_3). \eqn

Now we are interested in studying how these restrictions affect
the phase term. In general we have \eq {\bf p}\wedge {\bf q} =
\Theta(p_x q_y-p_y q_x). \eqn By choosing polar coordinates with
angle $\phi$, it follows that \eq {\bf p} \wedge {\bf q} = \Theta pq\left[
\cos(\phi_p)\sin(\phi_q)-\cos(\phi_q)\sin(\phi_p)\right] = \Theta
pq \sin(\phi_q-\phi_p). \eqn Then the phase is written as: \eq
\half\left[e^{-\frac{i\Theta }{ 2}( P_1 P_4 \sin(\phi_4-\phi_1) +
P_2 P_3 \sin(\phi_3-\phi_2) ) } + e^{-\frac{i\Theta }{ 2}( P_2
P_4\sin(\phi_4-\phi_2) + P_1 P_3 \sin(\phi_3-\phi_1) ) }\right].
\eqn

This, combined with the conditions for the angles, can be
rewritten as \eq F_\Theta(\phi_1;\phi_2) =
V(\phi_4=\phi_1;\phi_3=\phi_2;\phi_2;\phi_1)\half\left[1+
e^{-{i\Theta \over 2}(P_2 P_4 \sin(\phi_1-\phi_2) + P_1 P_3 \sin
(\phi_2-\phi_1))}\right], \eqn \eq F_\Theta'(\phi_1;\phi_2) =
V(\phi_4=\phi_2;\phi_3=\phi_1;\phi_2;\phi_1)\half\left[1+
e^{-{i\Theta \over 2}(P_1 P_4 \sin(\phi_2-\phi_1) + P_2 P_3 \sin
(\phi_1-\phi_2))}\right], \eqn
\begin{eqnarray}
V_\Theta(\phi_1;\phi_3)&=& V(\phi_4=-\phi_3;\phi_3;\phi_2=-\phi_1;\phi_1)
\half \left[
e^{-\frac{i\Theta}{2}(P_1 P_4 \sin(\phi_3-\phi_1+\pi)
+ P_2 P_3 \sin(\phi_3-\phi_1+\pi))} \right. \nonumber \\
&+& \left. e^{-\frac{i\Theta }{ 2}(P_2 P_4\sin(\phi_3-\phi_1) +
P_1 P_3 \sin(\phi_3-\phi_1))}\right],
\end{eqnarray}
or in a shorter form \eq F_\Theta(\phi_1;\phi_2) =
V(\phi_4=\phi_1;\phi_3=\phi_2;\phi_2;\phi_1)\half \left[1+ e^{-
{i\Theta \over 2}[(P_2 P_4-P_1P_3)\sin(\phi_1-\phi_2) ]}\right],
\eqn \eq F_\Theta'(\phi_1;\phi_2) =
V(\phi_4=\phi_2;\phi_3=\phi_1;\phi_2;\phi_1)\half\left[ 1+e^{-
{i\Theta \over 2}[(P_1 P_4 - P_2 P_3)
\sin(\phi_2-\phi_1)]}\right], \eqn
$$
V_\Theta(\phi_1;\phi_3) = V(\phi_4=\phi_3+\pi;\phi_2=\phi_1+\pi)
$$
\eq \times \half \left[e^{-\frac{i\Theta }{ 2}[(P_2 P_3+P_1 P_4
)\sin(\phi_1-\phi_3)]}+ e^{-\frac{i\Theta }{ 2}[(P_2 P_4+P_1
P_3)\sin(\phi_3-\phi_1)]} \right]. \eqn

We notice that in the first two expressions the original
antisymmetry of the interaction potential is lost; nevertheless
when we interchange, say, $P_1$ and $P_2$ the first expression
passes to the second. So the Fermi statistics is preserved. In the
third function interchanging two momenta is equivalent to adding
an angle $\pi$ to, say, $\phi_1$, and the phase term is invariant.

\section{One-Loop Renormalization of the Interaction}

In the previous section we have seen that we can write the
interaction term in the same form as in the usual commutative
case, absorbing the phase factors in the $ U_\Theta$ function;
then we can expand perturbatively as usual, resulting in the same
type of diagrams. The difference is that now we have extra
interesting phase factors.

For consistency we must compute to second order in $U_\Theta$,
which is equivalent to working at second order in the cummulant
expansion: \eq \half[\langle(\delta S)^2\rangle-\langle\delta
S\rangle^2]. \eqn

\begin{figure}[b]
\begin{center}
\includegraphics[height=4cm]{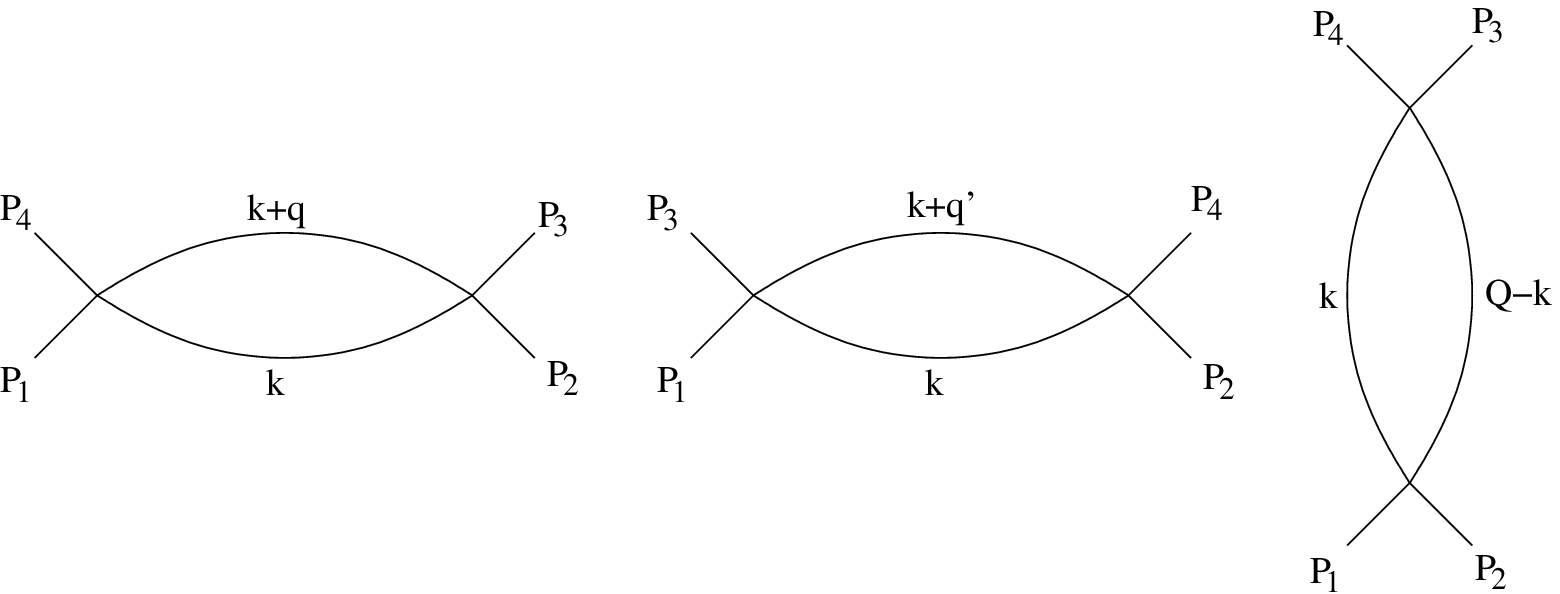}
\caption{Diagrams contributing to one-loop in the four-point
function.}\label{figdiagra}
\end{center}
\end{figure}

All disconnected diagrams are cancelled bf y the term $\langle\delta
S\rangle^2$, and the diagrams having non-vanishing contribution
are those shown in Fig. \ref{figdiagra}. The analytic expressions
are
\begin{eqnarray}
d\, U_\Theta&=&\int_\infty^\infty\int_{d\Lambda} {d\omega dK \over
4\pi^2}\int_0^{2\pi}{ d\phi \over 2\pi} {
U_\Theta({\bf P}_4,{\bf K}+{\bf q},{\bf K},{\bf P}_1)U_\Theta({\bf K},{\bf P}_3,{\bf P}_2,{\bf K}+{\bf q}) \over
[i\omega - E({\bf K})][i\omega -
E({\bf K}+{\bf q})]} \nonumber \\[2ex]
&&-\int_\infty^\infty\int_{d\Lambda} {d\omega dK \over
4\pi^2}\int_0^{2\pi} {d\phi \over 2\pi} {
U_\Theta({\bf P}_3,{\bf K}+{\bf q'},{\bf K},{\bf P}_1) U_\Theta({\bf K},{\bf P}_4,{\bf P}_2,{\bf K}+{\bf q'})
\over [i\omega -E({\bf K})][i\omega -
E({\bf K}+{\bf q'})]}\nonumber \\[2ex]
&&-{1 \over 2}\int_\infty^\infty\int_{d\Lambda} {d\omega dK \over
4\pi^2}\int_0^{2\pi} {d\phi \over 2\pi} {
U_\Theta({\bf Q}-{\bf K},{\bf K},{\bf P}_2,{\bf P}_1) U_\Theta({\bf P}_4,{\bf P}_3,{\bf Q}-{\bf K},{\bf K}) \over
[i\omega -E({\bf K})][-i\omega - E({\bf Q}-{\bf K})]},
\end{eqnarray}
where ${\bf q}={\bf P}_1-{\bf P}_4,$ ${\bf q'}={\bf P}_1-{\bf P}_3$ and $
{\bf Q}={\bf P}_1+{\bf P}_2$. The subscript $d\Lambda$ of the integral
indicates that both loop momenta in the diagram must be in the
thin shell being integrated. One of the internal line carries
momentum ${\bf K}$, which is restricted to the region defined bf y
$\Lambda$ around the Fermi surface. Implicitly the momentum of the
other internal line is ${\bf K}+{\bf q}$ in the ZS channel, ${\bf K}+{\bf q'}$ in
the ZS' channel, and ${\bf Q}-{\bf K}$ in the BCS channel, respectively.
We also impose the same conditions for the momentum variables
${\bf P}_1,{\bf P}_2,{\bf P}_3,{\bf P}_4$ to survive the renormalization at tree
level.

Before discussing the contribution of each diagram, let us pay
attention to the phase factors in these integrals, since they
contain the effects of space noncommutativity. We notice that the
phase term in the first integral is reduced to \eq \half
\left\{\cos\left[\frac{{\bf P}_1\wedge {\bf P}_4+ {\bf P}_2\wedge
{\bf P}_3}{2}\right]+ \cos\left[\frac{{\bf P}_1\wedge {\bf P}_4-{\bf P}_2\wedge
{\bf P}_3 +2{\bf K}\wedge({\bf P}_1-{\bf P}_4)}{2}\right]\right\}. \eqn
Analogously for the second integral we have \eq \half
\left\{\cos\left[\frac{{\bf P}_1\wedge {\bf P}_3+ {\bf P}_2\wedge
{\bf P}_4}{2}\right]+ \cos\left[\frac{{\bf P}_1\wedge {\bf P}_3-{\bf P}_2\wedge
{\bf P}_4 +2{\bf K}\wedge({\bf P}_4-{\bf P}_2)}{2}\right]\right\}. \eqn And
finally for the third integral the phase factor becomes \eq
\half\left\{\cos[{\bf K}\wedge({\bf P}_4-{\bf P}_1) + \half
{\bf Q}\wedge({\bf P}_1-{\bf P}_4)] + \cos[{\bf K}\wedge({\bf P}_4-{\bf P}_2) + \half
{\bf Q}\wedge({\bf P}_2-{\bf P}_4)]\right\}. \eqn

\subsection{Case I}

In this subsection we will analyze the first case (see Eq.
(\ref{cond1})) that survives the renormalization group analysis,
for the previous three channels: the ZS, the ZS' and the BCS ones,
respectively, in the noncommutative theory. We can see that in all
diagrams we have planar and non-planar contributions, then we need
to make a careful analysis of each diagram under the
renormalization conditions obtained at tree level.

For the condition that defines the function  $F$ (\ref{cond1}),
the phase factor in the first integral ( or the ZS channel) is \eq
\cos\left[\frac{(P_1P_4-P_2P_3)\Theta\sin(\phi_2-\phi_1)}{2}\right]
+ \cos\left[\frac{(P_1P_4+P_2P_3)\Theta\sin(\phi_2-\phi_1)}{2} +
\Theta Kq \sin(\phi_q-\phi)\right]. \eqn But as
$(P_1-P_4){bf \Omega}_1\approx 0$ this factor is reduced to \eq
1+\cos[ \theta K q\sin(\phi_q-\phi)], \eqn and considering that
${\bf q}={\bf P}_1-{\bf P}_4\approx 0$, this diagram is reduced to the usual
commutative one.

The phase factor for the ZS' diagram is \eq
\frac{1}{2}\left\{\cos\left[\frac{{\bf P}_1\wedge {\bf P}_3+ {\bf P}_2\wedge
{\bf P}_4}{2}\right]+ \cos\left[\frac{{\bf P}_1\wedge {\bf P}_3-{\bf P}_2\wedge
{\bf P}_4 +2{\bf K}\wedge({\bf P}_4-{\bf P}_2)}{2}\right]\right\}. \eqn With the
condition  (\ref{cond1}) this factor is reduced to \eq
\frac{1}{2}\left\{1+\cos[P_1P_3\Theta\sin(\phi_1-\phi_3)
+Kq'\Theta\sin(\phi_q-\phi)]\right\}. \eqn In this case one has
$q'\approx k_F$, thus the integral over the angle must be
restricted to an interval $d\Lambda/k_F$. The integration is \eq
\frac{1}{2}\int\frac{d\omega}{2\pi}\int_{d\Lambda}\frac{dK}{2\pi}
\int{d\phi}{2\pi}\frac{V(\phi_1,\phi)V(\phi,\phi_3)
[1+\cos[a+Kq'\Theta\sin(\phi_q-\phi)]]}{[i\omega-E(K)][i\omega-E(K+q')]},
\eqn where $a=P_1P_3\Theta\sin(\phi_1-\phi_3)$. This integral have
two contributions: The first one is planar, while the second one
needs a careful study. As $q'\approx k_F$, the poles of $\omega$
are in different half-planes, then this integral becomes \eq
\frac{1}{2}\int\frac{d\omega}{2\pi}\int_{d\Lambda}
\frac{dK}{2\pi}\int{d\phi}{2\pi}\frac{V(\phi_1,\phi)V(\phi,\phi_3)
[1+\cos[a+Kq'\Theta\sin(\phi_q-\phi)]]}{E(K)-E(K+q')}. \eqn Now
observe that $K$ is within an interval $d\Lambda$ around
$\Lambda$, so we can expand the cosine in series for $K$ around
$\Lambda$, then we get
\begin{eqnarray}\label{expan11}
\cos (a+2Kq'\Theta\sin(\phi_q-\phi))&\approx&
\cos (a+q'\Theta\Lambda\sin(\phi_q-\phi)) \nonumber \\
&-&q'\Theta(\Lambda-K)\sin(\phi_q-\phi)\sin(a+q'\Theta\Lambda
\sin(\phi_q-\phi))+ {\cal O}(\Theta^2).
\nonumber \\
\end{eqnarray}
With this expansion, the integral over $K$  from the first term in
the expansion gives us a term of order $d\Lambda$ and the $\omega$
integral gives a denominator of order $\Lambda$ due to the
restriction of the angle to  the range $d\Lambda/k_F$. Thus this
integral is of order $(d\Lambda/\Lambda)(d\Lambda/k_F)$; and the
$\beta$-function vanishes in the limit $|d\Lambda|/\Lambda
\rightarrow 0$.

The next term in the expansion is proportional to $\Theta$;
nevertheless, this term gives a contribution proportional to
$d\Lambda$ after integration over $K$,  so the contribution to
$\beta$-function is marginal. This conclusion is also valid for
higher order terms in $\Theta$.

For the BCS diagram, after using the condition (\ref{cond1}), one
can easily see that the phase factor is of the form \eq
\frac{1}{2}\left[1+\cos\left(K\wedge q'-\frac{1}{2}Q\wedge
q'\right)\right]. \eqn

This have the same form as the ZS' diagram, and the integral
limits are similar. So the contribution to  the $\beta$-function
vanishes also in the limit $d\Lambda/\Lambda\rightarrow 0$.

The analysis for the function $F_\Theta$ indicates that for the
case I, the space noncommtutativity does not induce any relevant
corrections and $F$ is a fixed point to this order.

\subsection{Case II}

For the case II (see the condition (\ref{cond2})), we have the
situation similar to that for the case I, in view of interchanging
${\bf P}_4\leftrightarrow {\bf P}_3 $. This is expected, because the
function $F_\Theta'$ allows one to recover the Fermi statistics.

\subsection{Case III}
In this case we take into account

 the condition III (\ref{cond3})
for each diagram, then for the ZS diagram the phase factor becomes
\eq \frac{1}{2}\left\{\cos
[P_1P_4\Theta\sin(\phi_4-\phi_1)+\cos[Kq\Theta\sin(\phi_q-\phi)
]\right\}, \eqn and the phase factor for the ZS' diagram is given
bf y \eq \frac{1}{2}\left\{\cos
[P_1P_3\Theta\sin(\phi_4-\phi_1]+\cos[Kq'\Theta\sin(\phi_q-\phi)
]\right\}. \eqn

For these diagrams the integral in $\omega$ gives a denominator of
order $\Lambda$, and the cosine function in the numerator can be
expanded as above, then the contribution to $\beta$-function
vanishes.

However, for the BCS diagram the phase factor is found to be \eq
\frac{1}{2}\left\{\cos[q\wedge K+\frac{1}{2}Q\wedge
q]+\cos[q'\wedge K+\frac{1}{2}Q\wedge q']\right\}. \eqn For this
diagram the angle is not restricted, so it can take any value.
Also in this diagram the integration over $\omega$ gives a
denominator of order $\Lambda$. Then we focus on the integration
over $K$.
Let us call the phase factor as ${\cal P}$. The integral to
calculate is \eq\label{chida}
-\frac{1}{2}\int_0^{2\pi}\frac{d\phi}{2\pi}\int_{d\Lambda}
\frac{dK}{2\pi}\frac{V(\phi_1-\phi)V(\phi-\phi_3)}{E(K)+E(Q-K)}{\cal
P}. \eqn We note that $K$ is in the region around Fermi surface
and so is $Q-K$. Thus we can take the approximation $E(K)\approx
vK$ and $E(Q-K)\approx v(Q-K)$, and therefore $E(K)+E(Q-K)\approx
vQ$ which is of order $\Lambda$. Then (\ref{chida}) gives
\eq\label{chidaone}
-\frac{1}{4\Lambda}\int_0^{2\pi}\frac{d\phi}{2\pi}\int_{d\Lambda}
\frac{dK}{2\pi}V(\phi_1-\phi)V(\phi-\phi_3)
\{\cos(qK\Theta\sin(\phi-\phi_q))
+\cos(q'K\Theta\sin(\phi-\phi_{q'}))\}. \eqn
As in the earlier cases we expand the cosine function around
$\Lambda$ so we have a similar expression as (\ref{expan11}) for
each term in the phase factor. We will consider only the first
terms up to order $\Theta$. Now the integration in $K$ becomes
easy and, for the same reason as in the previous cases, the terms
proportional to $(d\Lambda)^2$ can be neglected, because this kind
of terms make the $\beta$-function vanish in the limit
$d\Lambda/\Lambda\rightarrow 0$. After we make the change
$d\Lambda/\Lambda=dt$, we have a usual commutative contribution
with a factor of one half, plus the modifications due to
noncommutaivity; that is
\begin{eqnarray}\label{chida1}
\frac{dV(\phi_1-\phi_3)}{dt}&=&-\frac{1}{2}\int_0^{2\pi}
\frac{d\phi}{8\pi^2} V(\phi_1-\phi)V(\phi-\phi_3)
\{\cos[q\Lambda\Theta\sin(\phi-\phi_q)]\nonumber \\
&& \:\:-q\Theta\Lambda\sin(\phi-\phi_q)\sin[q\Lambda\Theta
\sin(\phi-\phi_{q})] + (q\leftrightarrow q')\}.
\end{eqnarray}

At this point it is convenient to express the functions $V(\phi)$
in terms of their Fourier components (or angular-momentum modes):
\eq V(\phi)=\sum_{l=0}^{\infty} V_le^{-il\phi}, \qquad {\rm
where}\qquad V_l=\int_0^{2\pi}\frac{d\phi}{2\pi}e^{il\phi}V(\phi).
\eqn This finally leads to a flow equation for the angular
momentum modes of $V$, in which different modes are coupled. After
integration we have a renormalization flow equation \eq
\sum_{l=0}^{\infty}\frac{dV_l}{dt}e^{-il(\phi_1-\phi_3)}
=-\frac{1}{8\pi}\sum_{l,l'=0}^{\infty} V_l V_{l'}
e^{-i(l\phi_1-l'\phi_3)} e^{i(l'-l)\phi_q}\left[ \frac{ d }{d\,
x_\Theta }\, x_\Theta J_{l'-l}(x_\Theta)   \right]  +
(q\leftrightarrow q'), \eqn where $x_\Theta=q\Lambda\Theta$. We
see that in this renormalization flow equation, there are highly
non-trivial contributions from space noncommutativity, that affect
the behavior of the BCS instability.

\section{Concluding Remarks}

In this paper we have used the renormalization group approach to
study how noncommutativity of spatial coordinates affects the low
energy behavior of a system of weakly interacting fermions. The
physics of the Gaussian fixed point still corresponds to the
Landau theory, whose excitations are still the Landau
quasiparticles, not the bare particles, as in the ordinary
Fermi-Landau liquids. But the properties of Landau quasiparticles
gets modified, in consistency with the UV-IR mixing, a general
feature of noncommutative field theory. In particular, we found
that at one loop level, the pairing instability in the BCS channel
(say in $d=2$ cases) gets modified through the noncommutative
corrections to the flow equation for the interaction function
$V_\Theta$.

In our study we have considered the simplest case with a circular
Fermi surface in two spatial dimensions. It would be worth to
analyze the more general cases. Also we have concentrated on
low-energy phenomena happening near the Fermi surface. We expect
that working away from the Fermi surface, one could have some new
nontrivial contributions from the noncommutative parameter. Some
work in this regard is in progress.

Finally, we would like to make the remark that in our analysis, we
are interested only in the one loop noncommuative corrections
appearing in the low energy regime of the weakly interacting
theory. At this level the theory is stable, as we can see from the
fact that there are no corrections due to the noncommutativity in
self energy. Non-planar corrections in the BCS channel is expected
to contribute to the two point function at two loop, but going to
higher loops is beyond the scope of the present paper.

\vskip 1truecm
\centerline{\bf Acknowledgments}

The work of S.E.-J. was supported by a CONACyT postdoctoral
fellowship and Grant PROMEP /103.5/08/3291. The work of H.G.-C.
was supported in part by CONACyT M\'exico Grant 45713-F. YSW was supported in part by US NSF through Grant No. PHY-0756958.


\bibliography{octaviostrings}
\addcontentsline{toc}{section}{Bibliography}
\bibliographystyle{TitleAndArxiv}


\begin{figure}[b!]
\begin{center}
\includegraphics[height=4cm]{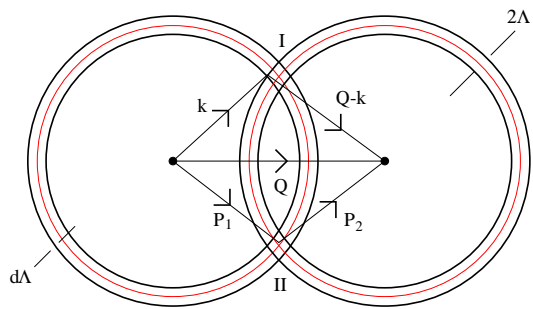}
\caption{Configuration of momenta near Fermi surface with shell
width $2\Lambda$.}\label{figconf}
\end{center}
\end{figure}
\end{document}